\documentclass[12pt]{article}
\usepackage{amssymb}
\usepackage{epsfig}

\catcode`\@=11
\def\numberbysection{\@addtoreset{equation}{section}
        \def\theequation{\thesection.\arabic{equation}}}

\def\beq{\begin{equation}}
\def\eeq{\end{equation}}
\numberbysection
\begin{document}
\begin{titlepage}
\begin{center}
\hfill  \\
\vskip 1.in {\Large \bf Harmonic oscillators in a Snyder geometry} \vskip 0.5in P. Valtancoli
\\[.2in]
{\em Dipartimento di Fisica, Polo Scientifico Universit\'a di Firenze \\
and INFN, Sezione di Firenze (Italy)\\
Via G. Sansone 1, 50019 Sesto Fiorentino, Italy}
\end{center}
\vskip .5in
\begin{abstract}
We find that, in presence of the Snyder geometry, the quantization of d isotropic harmonic oscillators can be solved exactly.
\end{abstract}
\medskip
\end{titlepage}
\pagenumbering{arabic}
\section{Introduction}

Recent studies have suggested that a natural cutoff for the ultraviolet behaviour of the physical theories can be obtained modifying the Heisenberg algebra of the canonical commutation rules \cite{1}-\cite{2}-\cite{3}-\cite{4}-\cite{5}. This implies a finite minimal uncertainty $\Delta x_0$ in the position measurement, modifying the structure of space-time at short distances.

Such cutoff can be already applied in the non-relativistic approximation, i.e. in ordinary quantum mechanics. The most elementary example is the harmonic oscillator with a finite
minimal uncertainty $\Delta x_0$, a problem solved in \cite{1}, using the momentum representation.

It is the aim of this article the generalization of this result to the case of $d$ isotropic harmonic oscillators quantized with the Snyder algebra. We will show that this problem can be exactly solved, without any approximation, allowing us to discuss how the state degeneracy of $d$ independent harmonic oscillators is removed by the non-commutative deformation.

To reach such aim we have studied the eigenvalue equation in a new representation, the variables $\rho_i$ which resolve the Snyder algebra. As an outcome of our research, we have been able to solve, at a mathematic level, a $d$-dimensional generalization of the well known Gegenbauer equation.

\section{Harmonic oscillator revisited}

In \cite{1} the harmonic oscillator has been quantized using the following modified quantization rule:

\beq [ \ \hat{x} \ , \ \hat{p} \ ] \ = \ i \hbar \ ( \ 1 \ + \ \beta \ \hat{p}^2 \ ) \label{21} \eeq

in the momentum representation. The corresponding eigenvalue equation is

\beq ( \ \hat{p}^2 \ + \ m^2 \omega^2 \hat{x}^2 \ ) \ \psi(p) \ = \ 2 m E \ \psi(p) \label{22} \eeq

The idea behind our present paper is to solve this problem in a new representation \cite{2}

\begin{eqnarray}
\hat{x} & = & \ i \hbar \ \sqrt{ 1 \ - \ \beta \rho^2 } \ \frac{\partial}{\partial \rho} \ \ \ \ \ \ \ \ \ \ \ \ \ \  0 < \rho < \frac{1}{\sqrt{\beta}} \nonumber \\
\hat{p} & = & \frac{\rho}{\sqrt{ 1 \ - \ \beta \rho^2 }} \label{23}
\end{eqnarray}

This effort will be helpful in the next section to solve exactly the quantization of $d$ harmonic oscillators in the Snyder geometry.

The eigenvalue equation in the variables $\rho$ turns out to be

\beq \left( \frac{\rho^2}{1 \ - \ \beta \rho^2} \ - \ \frac{1}{\beta^2 \ \mu  (  \mu  -  1 )} \ \sqrt{  1  - \beta  \rho^2  } \ \frac{\partial}{\partial \rho} \
\sqrt{  1  - \beta  \rho^2  } \ \frac{\partial}{\partial \rho} \ \right) \psi(\rho) \ = \ 2  m  E  \ \psi(\rho) \label{24}\eeq

where $ m \omega \hbar \beta \ = \ {[ \ \mu  (  \mu  -  1  ) \ ] }^{-\frac{1}{2}} $.

For the ground state it is well known that the solution is

\beq \psi_0 ( \rho ) \ = \ c_0 \ {(1-\beta\rho^2)}^{\frac{\mu}{2}} \label{25} \eeq

with eigenvalue

\beq E_0 \ = \ \frac{\hbar \omega}{2} \ \sqrt{\frac{\mu}{\mu-1}} \label{26} \eeq

To study the excited states we look for a solution of this type

\beq \psi ( \rho ) \ \sim \ \chi ( \rho ) \ {(1-\beta\rho^2)}^{\frac{\mu}{2}} \label{27} \eeq

from which we obtain the following differential equation for $ \chi ( \rho ) $

\beq {(1-\beta\rho^2)} \ \frac{\partial^2}{\partial \rho^2} \ \chi(\rho) \ - \ \beta ( 2\mu+1) \ \rho \ \frac{\partial}{\partial\rho} \ \chi(\rho) \ + \ \beta \mu \ [ 2 m E \beta ( \mu-1) - 1 ] \ \chi( \rho ) \ = \ 0 \label{28} \eeq

Let us make the substitution

\beq E \ = \ \frac{\hbar \omega}{2} \ \frac{ \nu^2 + ( 2\nu + 1 ) \mu }{ \sqrt{ \mu ( \mu-1 )}} \label{29} \eeq

and change the variable $ \rho \rightarrow z = \sqrt{\beta} \  \rho $, obtaining

\beq ( 1-z^2 ) \ \frac{\partial^2}{\partial z^2} \ \chi(z) \ - \ ( 2\mu + 1 ) \ z \ \frac{\partial}{\partial z} \ \chi(z) \ + \ \nu ( \nu + 2 \mu ) \ \chi(z) \ = \ 0 \label{210} \eeq

We easily recognize the Gegenbauer equation, whose polynomial solutions are obtained for

\beq \nu \ = \ n \ \ \ \ \ \Rightarrow \ \ \ \ \ E_n \ = \ \frac{\hbar \omega}{2} \ \frac{ n^2 + ( 2 n + 1 ) \mu }{ \sqrt{ \mu ( \mu-1 )}} \label{211} \eeq

The corresponding eigenfunctions are proportional to the Gegenbauer polynomials, satisfying the recurrence equation

\beq \partial_z \ P^\mu_n (z) \ = \ 2 \mu \ P^{\mu+1}_{n-1} (z) \label{212} \eeq

Similarly to the Hermite polynomials, there is a generating function for the Gegenbauer polynomials

\beq \frac{1}{ ( 1 - 2 z t + t^2 )^\mu } \ = \ \sum^{\infty}_{n=0} \ P^\mu_n (z) t^n \label{213} \eeq

The first polynomials are

\begin{eqnarray}
P_0^\mu (z) & = & 1 \nonumber \\
P_1^\mu (z) & = & 2 \mu z \nonumber \\
P_2^\mu (z) & = & - \mu + 2 \mu ( 1+\mu ) z^2 \nonumber \\
P_3^\mu (z) & = & - 2 \mu ( 1+\mu ) z + \frac{4}{3} \mu ( 1+\mu ) ( 2+\mu ) z^3
\label{214}
\end{eqnarray}

\section{Harmonic oscillators in d dimensions}

The theory of the harmonic oscillator described in the previous section can be generalized in $d$ dimensions. The Hamiltonian of isotropic oscillators of equal mass $m$ and frequency $\omega$ in $d$ dimensions and cartesian coordinates can be written as

\beq H \ = \ \sum^{d}_{i=1} \ \left( \frac{\hat{p}^2_i}{2m} \ + \ \frac{1}{2} m \omega^2 \hat{x}^2_i \right) \label{31} \eeq

For the generic $d$ case, we extend the commutation rule (\ref{21}) to the Snyder algebra \cite{3}

\begin{eqnarray}
& \ & [ x_i, p_j ] \ = \ i \hbar \  ( \delta_{ij} \ + \ \beta p_i p_j ) \nonumber \\
& \ & [ p_i, p_j ] \ = \ 0 \nonumber \\
& \ & [ x_i, x_j ] \ = \ i \hbar \beta \ ( p_j x_i \ - \ p_i x_j ) \label{32}
\end{eqnarray}

This algebra is resolved by the $\rho$ representation \cite{2}

\beq x_i \ = \ i \hbar \ \sqrt{ 1 -\beta\rho^2 } \ \frac{\partial}{\partial\rho_i} \ \ \ \ \ \ \ \ \ \ \ p_i \ = \ \frac{\rho_i}{\sqrt{ 1-\beta\rho^2 }} \ \ \ \ \ \ \ \
0 < \rho^2 < \frac{1}{\beta}
\label{33} \eeq

The eigenvalue equation for $d$ oscillators with the same frequency $\omega$ and mass $m$ is the following:

\beq \sum_{i=1}^d \ ( \hat{p}^2_i \ + \ m^2 \omega^2 \hat{x}^2_i ) \ \psi( \rho ) \ = \
2 m E \ \psi ( \rho ) \label{34} \eeq

that, rewritten in the $\rho$ variables, reads

\beq \sum^d_{i=1} \ \left( \frac{\rho^2_i}{1-\beta\rho^2} \ - \ \frac{1}{\beta^2 \mu ( \mu-1 )} \ \sqrt{ 1-\beta\rho^2 } \ \frac{ \partial }{ \partial \rho_i } \ \sqrt{ 1-\beta\rho^2 } \ \frac{ \partial }{ \partial \rho_i }
\right) \ \psi( \rho ) \ = \ 2 m E \ \psi ( \rho ) \label{35} \eeq

The ground state is simply

\beq \psi_0 ( \rho ) \ = \ c_0 \ {( 1-\beta\rho^2 )}^{\frac{\mu}{2}} \label{36} \eeq

with eigenvalue

\beq E_0 \ = \ d \ \frac{\hbar \omega}{2} \ \sqrt{\frac{\mu}{\mu-1}} \label{37} \eeq

To study the excited states, we introduce as in the case $d=1$ the following ansatz

\beq \psi ( \rho ) \ = \ \chi ( \rho ) \ {( 1-\beta\rho^2 )}^{\frac{\mu}{2}} \label{38} \eeq

from which we obtain

\beq ( 1-\beta\rho^2 ) \ \sum^d_{i=1} \ \frac{\partial^2}{\partial\rho_i \partial\rho_i} \ \chi( \rho ) \ - \ \beta ( 1 + 2 \mu ) \sum^d_{i=1} \ \rho_i \ \frac{\partial}{\partial \rho_i} \ \chi ( \rho ) \ + \ \beta \mu \ [ \ 2 m E  \beta ( \mu-1 ) \ - \ d \ ] \ \chi ( \rho ) \ = \ 0 \label{39} \eeq

a $d$-dimensional generalization of the Gegenbauer equation.

Let us introduce the notation $\epsilon_{\{n_i\}} \ = \ \mu \ [ \ 2 m E \beta ( \mu-1 ) \ - \ d \ ]$ and  $z_i \ = \ \sqrt{ \beta } \rho_i$, where the parameter $\epsilon$ depends
on some quantum numbers $n_i$, from which the equation to be solved is

 \beq \left[ \ [ 1 - ( \sum^d_{i=1} z^2_i ) ] \ \sum^d_{i=1} \ \frac{\partial^2}{\partial z_i \partial z_i} \ - \ ( 1+2\mu ) \ \sum^d_{i=1} z_i \frac{\partial}{\partial z_i} \ + \ \epsilon_{\{n_i\}} \ \right] \ P_{\{ n_i \}} ( z_i ) \ = \ 0 \label{310} \eeq

with the energy parameter given by

 \beq E_{\{n_i\}} \ = \ \frac{\hbar \omega}{2} \ \frac{ ( \ \epsilon_{\{n_i\}} \ + \ d \  \mu \ ) }{\sqrt{\mu ( \mu-1 )}} \label{311} \eeq

The symmetry of this equation suggests to introduce the following ansatz, i.e. that the polynomial solution space is composed by two parts:

i) the solutions $P_{kk}$  of the free differential equation

\beq \left( \ \sum^d_{i=1} \ \frac{\partial^2}{\partial z_i \partial z_i} \ \right) \ P_{kk} ( z_i ) \ = \ 0
\label{312}
 \eeq

where $P_{kk} ( z_i )$  is an homogeneous polynomial in the variables $z_i$ of degree $k$.

In this case it is easy to compute the corresponding energy eigenvalue

\beq \epsilon_{kk}^\mu \ = \ k ( 1+ 2\mu ) \label{313} \eeq

ii) the solutions $P_{N k}$ with $ N = k + 2n $ where

\beq P_{N k} ( z_i ) \ = \ P_{kk} ( z_i ) \left( \ 1 \ + \ \sum^n_{i=1} a_i \ \left( \sum^d_{j=1} z^2_j \right)^i \ \right) \label{314} \eeq

To compute the eigenvalue corresponding to $ P_{Nk} ( z_i ) $ we apply the following differential operator $ \sum^d_{i=1} \ \frac{\partial^2}{\partial z_i \partial z_i} $
to the eigenvalue equation (\ref{310}). With simple algebraic steps we deduce that

\beq \left( \ \sum^d_{i=1} \ \frac{\partial^2}{\partial z_i \partial z_i} \ \right) \ P^\mu_{N k} \ ( z_i ) \ \propto \ P^{\mu+2}_{(N-2) k} ( z_i ) \label{315} \eeq

completed with the energy constraint

\begin{eqnarray}
& \ & \epsilon_{Nk}^\mu  \ = \ \epsilon_{(N-2) k}^{\mu+2}  \ + \ 2 ( 1+2\mu ) + 2d \nonumber \\
& \ & \epsilon_{kk}^{\mu} \ = \ k ( 1+2\mu ) \label{316} \end{eqnarray}

This recurrence equation (\ref{316}) is solved by

\beq \epsilon_{N k}^\mu \ = \ N ( 1+2\mu ) \ + \ ( N-k ) ( N + k + d-2 ) \label{317}
\eeq

from which we deduce that the energy eigenvalues depends only on two quantum numbers
$N$ and $k$ ( whose difference $ N-k $ must be an even positive integer number )

\beq E_{Nk} \ = \frac{\hbar \omega}{2} \ \frac{ [ N ( 1+ 2\mu ) \ + \ ( N-k ) ( N+k+d-2 ) + d \mu ]}{\sqrt{\mu ( \mu-1 )}} \label{318} \eeq

We conclude that, comparing with the case $\beta=0$, the noncommutative deformation (\ref{32})
reduces the states degeneracy from $d-1$ degrees of freedom to $d-2$.

Now it remains to show that the ansatz with which we have solved the differential equation (\ref{310}) gives all the polynomial solutions. We know that for $ \beta \rightarrow 0 $ our problem reduces to $d$ independent oscillators; in this case fixing the level
$N$, the number of independent eigenfunctions is given by the formula

\beq s_d (N) \ = \ \frac{ ( N+d-1)! }{ N! \ (d-1)! } \label{319} \eeq

For $\beta \neq 0$ we must simply count how many independent solutions $s_d (k)$ exist of the free differential equation (\ref{312}) defining  $ P_{kk} (z_i)$. We have computed them until  $d=5$ oscillators

\begin{eqnarray}
& \ & s_2 (k) \ = \ 2   \ \ \ \ \ \ \ \ \ k > 0 , \ \ \ \ \ \ \ s_2(0) \ = \ 1 \nonumber \\
& \ & s_3 (k) \ = \ 2k +1 \nonumber \\
& \ & s_4 (k) \ = \ ( k+1 )^2 \nonumber \\
& \ & s_5 (k) \ = \ \frac{k ( k+1 ) ( 2k+1)}{6} \ + \ ( k+1 )^2
\label{320} \end{eqnarray}

In all cases we can check that the following identities hold

\begin{eqnarray}
s_d ( N ) & = & \sum^{N}_{k = {\rm even}} \ s_d ( k ) \ \ \ \ \ \ \ \ \ \ N \ \ {\rm even} \nonumber \\
& = & \sum^{N}_{k = {\rm odd} } \ s_d (k ) \ \ \ \ \ \ \ \ \ \ \ \ \ N \ \ {\rm odd}
\label{321}
\end{eqnarray}

This completes our proof that we have described a complete basis of the Hilbert space.

\section{Conclusion}

In this article we have shown that, even modifying the quantization rule, many problems of quantum mechanics can be exactly solved. In particular we have found that the natural $d$-dimensional extension of the modified Heisenberg algebra (\ref{21}) is surely the Snyder algebra (\ref{32}). We have discussed in detail that the quantization of $d$ isotropic oscillators
in non-commutative geometry gives rise to two relevant quantum numbers, from which we can deduce that the residual degeneracy of the states is reduced to $d-2$ degrees of freedom. The spectrum contains, besides a linear term in the main quantum number $N$ ( that in the commutative limit is the sum of the single particle quantum numbers $n_i$ ),
a quadratic term dependent also on a secondary quantum number $k$, such as $N-k$ is an even positive integer number. In the limit $d \rightarrow 1$, our general formula reduces to the single harmonic oscillator spectrum studied in \cite{1}.

We therefore expect that the solvability of these examples can be extended to more complex cases.  Work is in progress in this direction.

\end{document}